\documentclass[twocolumn,9pt]{article}
\usepackage{extsizes}
\usepackage[super,sort&compress,comma]{natbib} 
\usepackage[version=3]{mhchem}
\usepackage[left=1.5cm, right=1.5cm, top=1.785cm, bottom=2.0cm]{geometry}
\usepackage{balance}
\usepackage{times,mathptmx}
\usepackage{sectsty}
\usepackage{graphicx} 
\usepackage{lastpage}
\usepackage[format=plain,justification=raggedright,singlelinecheck=false,font={stretch=1.125,small,sf},labelfont=bf,labelsep=space]{caption}
\usepackage{float}

\usepackage{fancyhdr}
\usepackage{fnpos}
\usepackage[english]{babel}
\usepackage{array}
\usepackage{droidsans}
\usepackage{charter}
\usepackage[T1]{fontenc}
\usepackage[usenames,dvipsnames]{xcolor}
\usepackage{setspace}
\usepackage[compact]{titlesec}

\usepackage{braket}

\definecolor{cream}{RGB}{222,217,201}
\title{Interference stabilization of autoionizing states in molecular N$_2$ studied by time- and angular-resolved photoelectron spectroscopy.} 
\author{Martin Eckstein,\textit{$^{a}$}
                   Nicola Mayer,\textit{$^{a}$}
				   Chung-Hsin Yang,\textit{$^{a}$}
				   Giuseppe Sansone,\textit{$^{b}$}
				   Marc J. J. Vrakking,\textit{$^{a}$}
				   Misha Ivanov,\textit{$^{a}$}\\
				   and Oleg Kornilov$^{\ast}$\textit{$^{a}$}\\
				   [1em]
				   $^{a}$~Max Born Institute, Max-Born-Stra{\ss}e 2A, 12489 Berlin, Germany. E-mail: kornilov@mbi-berlin.de\\ $^{b}$~Dipartimento di Fisica, Politecnico, Piazza Leonardo da Vinci 32, 20133 Milano, Italy. }
\date{}

\newcommand{\ress}{4`s'$\sigma_g$}
\newcommand{\resd}{3d$\pi_g$}
\begin{document}
\maketitle
\begin{abstract}An autoionizing resonance in molecular N$_2$ is excited by an ultrashort XUV pulse and probed by 
a subsequent weak IR pulse, which ionizes the contributing Rydberg states. Time- and angular-resolved photoelectron spectra recorded with 
a velocity map imaging spectrometer reveal two electronic contributions with different angular distributions. One of them has an exponential
decay rate of $20\pm5$~fs, while the other one is shorter than 10~fs. This observation is interpreted as a manifestation of interference stabilization involving
the two overlapping discrete Rydberg states. A formalism of interference stabilization for molecular ionization is developed and applied to
describe the autoionizing resonance. The results of calculations reveal, that the effect of the interference stabilization is facilitated by 
rotationally-induced couplings of electronic states with different symmetry.\end{abstract}

\section{Introduction}

One of the main challenges in quantum chemistry is the description of electron-electron correlations, which are at the root of electron dynamics. 
The recent development of experimental methods with time resolution comparable to the characteristic time of electron motion opens the 
possibility to study electron-electron correlations directly in the time domain. These methods have recently been applied to
autoionization in helium \cite{Ott2013,Ott2014}, neon \cite{Mashiko2014}, and argon atoms \cite{Kotur2016}, paving the way to 
studies of multielectron dynamics from the time domain perspective. All these studies are concerned 
with isolated autoionizing resonances, which are represented in the spectral domain as Fano line profiles \cite{Fano1961}. In our recent work we used 
time-, energy-, and angular-resolved photoelectron spectroscopy to investigate for the first time dynamics of autoionization at a complex resonance
in nitrogen molecules \cite{Eckstein2016}. In this article we expand on the previous report providing a detailed analysis of the angular distributions and 
developing a theoretical framework for the description of coupled molecular resonances based on the theory of interference stabilization 
\cite{Heller1976, Feneuille1982, Liu1985, Fedorov1988}. Our experimental and theoretical results for the first time demonstrate the
presence of this so-far elusive mechanism in molecular autoionization.

Autoionizing resonances in nitrogen molecules were first discovered in 1936 by Hopfield, who observed two sequences of sharp spectral lines
converging to a common threshold \cite{Hopfield1930}. It was soon realized that the lines are caused by the excitation of electronic Rydberg 
states attached to nitrogen ion in the $B^2\Sigma_u^+$ electronic state, which decays via electron rearrangement resulting in autoionization. However, 
it took until 1961 to explain the asymmetric line profiles, later called Fano profiles, which are caused by an interference 
between the direct ionization and the autoionization \cite{Fano1961}. Even today the details of the line profiles of the two series observed by 
Hopfield (the absorption and the ''emission'' series) have not been fully described theoretically. The most detailed calculations of Raoult et al 
\cite{Raoult1983} provided correct assignments of the resonance lines to the responsible Rydberg series, but could not reproduce the line 
shapes. Later experiments of Huber et al \cite{Huber1993} with room temperature nitrogen and rotationally-cold supersonic jets demonstrated, that 
rotational couplings of electronic states with different angular momenta are important. These coupling were not included in the calculations of Ref. 
\cite{Raoult1983}) carried out in the space-fixed molecular frame. 

\begin{figure*}[htbp!]
  \centering
    \includegraphics[width=\textwidth]{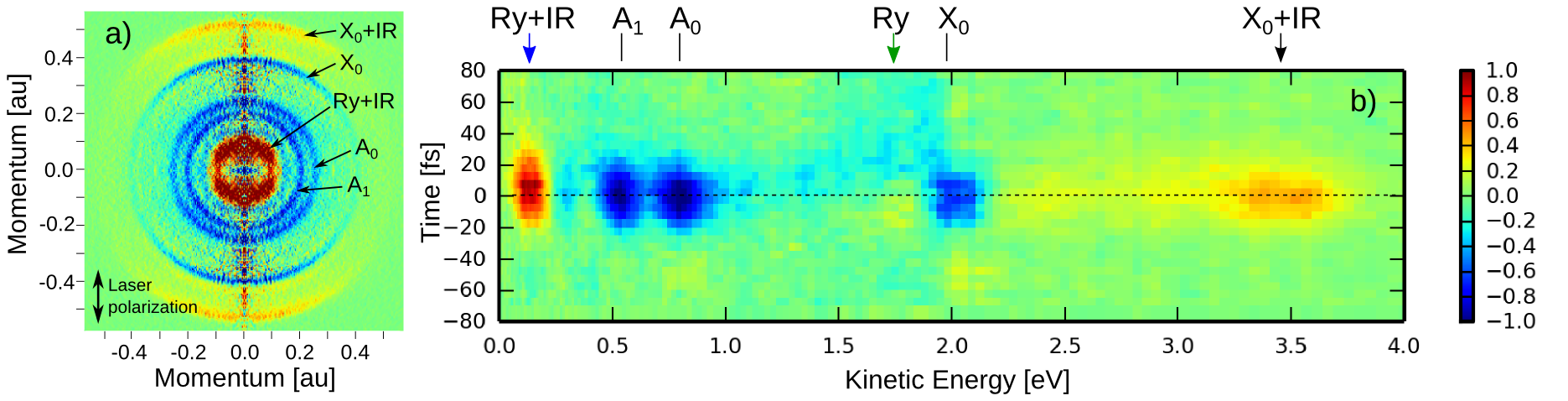}
      \caption{\label{fig:VMI} a) A difference VMI image at zero time delay between the XUV and the IR pulses. The image is obtained by 
	           subtracting an XUV-only signal from an XUV+IR one. The double arrow in the bottom left corner indicates  the XUV and IR laser 
			   polarization. Features corresponding to direct ionization to the $X^2\Sigma_g^+$ and $A^2\Pi_u$ channels are labeled as $X_0$, 
			   $A_{0,1,2}$, where the subscript denotes the vibrational state. The most intense IR-assisted sideband signal is labeled 
			   $X_0$+IR. 			   
			   The feature corresponding to IR ionization of the Rydberg state attached to the $B^2\Sigma_u^+$ ionization threshold is labeled as 
			   Ry+IR. b) False color map of photoelectron kinetic energy spectra plotted vs. kinetic energy and 
			   time delay between the XUV and IR pulses. The dashed line indicates the position of time ''zero'' determined from the X$_0$+IR sideband 
			   signal.	The map clearly shows that only the Ry+IR feature extends towards positive time delays. Other intense features are symmetric with respect to $t=0$~fs, i.e. show no dynamics. }
\end{figure*}

In this work the resonance at 17.33~eV is investigated. The results of Raoult et al \cite{Raoult1983} suggest that at the position of 
this member of the emission series two Rydberg states overlap, namely the {\resd} and {\ress} states.
This prediction is confirmed in our time-resolved experiments \cite{Eckstein2016}, which reveal time-dependent asymmetry parameters in the  
transient photoelectron signals pointing to contributions of at least two electronic states. The results further show that the two 
participating electronic states have different lifetimes contrary to the predictions of Ref. \cite{Raoult1983}. This discrepancy
suggests that the so-called phenomenon of interference stabilization may play a role, 

The phenomenon of interference stabilization, also called interference narrowing, was previously observed and described for excited atoms in
static electric fields and laser fields \cite{Heller1976, Feneuille1982, Liu1985, Stodolna2014}. It was described in detail analytically by 
Fedorov and Movsesian \cite{Fedorov1988} for a set of overlapping resonances in atomic ionization by a strong laser field. In their model a strong laser field ionizes
two or more Rydberg states, i.e. couples them to the ionization continuum. As the laser field strength increases, the lifetimes of individual Rydberg states decrease and hence their corresponding energy widths increase. When the widths of the 
states become comparable to the energy spacing between them, the continuum couplings (i.e. the interactions between the resonances
via the continuum) start to play a significant role and lead to a non-intuitive effect: the increase of their lifetimes. In case of two resonances, the energies of
the two states approach each other and the width of one of the resonances continues to grow with the laser field strength, while the width of the 
other state starts to shrink. That is, one of the resonances becomes progressively more and more stabilized, even though the coupling constants for
both discrete states continue to increase. This effect was termed ''interference stabilization'' by Fedorov et al \cite{Fedorov1988}, because it 
originates from quantum interference of the contributions to ionization resulting from the two states. Its origin is in general similar to that of the electromagnetically induced transparency (EIT). 

The origin of the stabilization effect is generic and its theory is applicable to all types of discrete-continuum couplings, not necessarily induced 
by static electric fields or laser fields. It is 
therefore reasonable to expect it to influence the case of two overlapping Fano resonances, where electron-electron interactions are responsible for 
coupling autoionizing Rydberg states to the continuum. In fact, such an effect was numerically predicted by Mies in 1968 \cite{Mies1968}, although he 
did not analyze it analytically. In this article we generalize the theory of Fedorov et al following the Ref. \cite{Fedorov1998} to the 
realistic case of autoionization of N$_2$, combining the treatment of Fano resonances with the interference stabilization effect,
 and analyze the results of time-resolved photoelectron spectroscopy experiments in terms of the theory of 
interference stabilization.

The articles is structured as follows: the next section briefly describes the experiment and presents results on time-dependent angular distributions
of the N$_2$ autoionizing resonance at 17.33~eV. The third section presents a theoretical description of the interference stabilization
effect for the case of N$_2$. The forth section provides a comparison of the theoretical and the experimental results.

\section{Experimental results}

The experiment has already been described in detail in our previous report \cite{Eckstein2016}. Here we briefly summarize the experimental setup, the
main findings and analyze in detail the time-dependent photoelectron angular distributions corresponding to the 17.33~eV resonance 
investigated in the present work. Time-resolved study of the Fano resonance in N$_2$ was performed at a recently constructed XUV time delay
compensating monochromator beamline connected to a velocity map imaging (VMI) spectrometer \cite{Eckstein2015}. In the experiment 2~mJ, 20~fs 
near-infrared (IR) pulses from a commercial Ti:sapphire laser system were used to drive a high-order harmonic
generation process in Ar gas \cite{Pfeifer2006} to produce linearly polarized XUV pulses with photon energies ranging from 10 to 70~eV. The time 
delay compensating monochromator \cite{Eckstein2015} was used to select the 11th harmonic of the fundamental wavelength and slightly tune 
its spectrum to predominantly excite the resonance at 17.33~eV. For this purpose the XUV excitation spectrum was centered at 17.5~eV and had a full width at 
half maximum (FWHM) of 0.5~eV.
Autoionizing Rydberg states excited by the XUV pulse were probed by a weak IR pulse with a pulse energy of 50~$\mu$J 
(intensity of 2~TW/cm$^2$) and a
linear polarization parallel to that of the XUV pulse. The pulse arrives with a variable time delay with respect to the XUV pulse. The resulting photoelectron 
spectra are recorded using the VMI spectrometer \cite{Ghafur2009}, which provides both energy- and angular-resolved photoelectron spectra. The changes 
induced by the IR pulse are modest and therefore the results are best represented in the form of difference VMI images, i.e. images resulting from 
subtraction of an XUV-only image from an image recorded when both the XUV and IR pulses are present. Such a difference image recorded at zero time delay 
between the XUV and IR pulses is presented in Fig. \ref{fig:VMI}a). The Figure shows a slice of the 3D momentum distribution 
reconstructed from the raw experimental data using the BASEX reconstruction algorithm \cite{Dribinski2002}. The image contains several features 
having both positive (red, additional signal due to the IR pulse) and negative (blue, signal reduced by the IR pulse) intensities. The outermost feature, labeled $X_0$+IR, corresponds to the generation of a $n=+1$ sideband
\cite{Kroll1973,Maquet2007} of the photoelectron feature in the $X^2\Sigma_g^+$($\nu=0$) ionization 
channel, i.e. the channel in which the nitrogen molecular ion N$_2^+$ is in the ground electronic and vibrational state. The negative (blue) feature 
labeled X$_0$ is a depletion of the signal in the $X^2\Sigma_g^+$($\nu=0$) channel due to formation of the sideband. Similar signals are observed 
for the $A^2\Pi_u$($\nu=0,1,2$) ionization channels, but the corresponding sideband signals are partially masked by the dominant $X_0$ feature.
The signal corresponding to ionization of the Rydberg electron by one IR photon appears as an intense positive (red) ring close to the center of the
image. It is labeled Ry+IR to reflect its nature: ionization of a Rydberg state with one IR photon leaving the core in the excited $B^2\Sigma_u$ 
state.

The temporal evolution of the features observed in the photoelectron spectra was studied by scanning the time delay between the XUV and IR pulses.
The false color map of the photoelectron kinetic energy spectra (integrated over all angles) as a function of the XUV-IR delay is shown in
Fig. \ref{fig:VMI}b). As expected, all signals corresponding to a formation of sidebands and depletion of the original photoelectron bands have 
temporal profiles symmetric with respect to the zero time delay (dashed line). On the contrary, feature Ry+IR extends towards positive 
time delays (where the XUV pulse arrives before the IR pulse), which reflects the finite lifetime of the Rydberg state. 

\begin{figure}[hbpt!]
  \centering
    \includegraphics[width=\columnwidth]{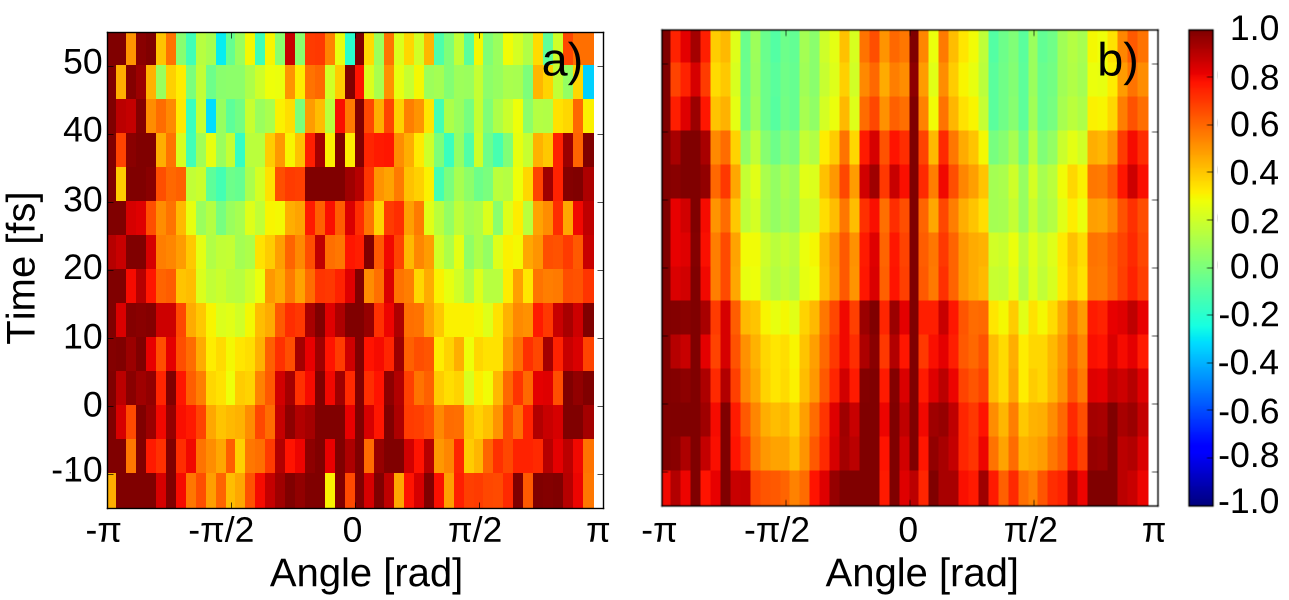}
      \caption{\label{fig:angmap} a) False color map of the photoelectron angular distributions of feature Ry+IR normalized to the intensity at zero 
	  emission angle (relative to the light polarization). This plot emphasizes the time-dependent changes of the angular distributions as the
	  signal becomes more peaked along the 0 and $\pm\pi$ directions for positive time delays. The data are plotted in the range of time delays,
          where the signal intensity exceeds the noise.
          b) False color map of the photoelectron angular distributions composed of the two dominant components
	  extracted from singular value decomposition of the experimental data matrix plotted in a). This procedure removes experimental noise and allows
	  for extraction of decay-associated angular distributions (see text for the details).
	  }
\end{figure}

The short lifetime of the resonance in Fig. \ref{fig:VMI}b) precludes an analysis of its composition based on
the transient intensity data alone. However, following and extending the approach of Ref. \cite{Eckstein2016} the time-dependent angular distributions
of this feature can be used to analyze its origin. Fig. \ref{fig:angmap}a) shows the angular distributions of feature Ry+IR integrated over
a range of energies from 0.1 to 0.3~eV as a function of time delay. The distributions are normalized to the maximum signal intensity at each
pump-probe delay. This normalization procedure removes the time dependence of the signal intensity and thus reveals pure variations of the angular 
distributions of the photoemission as a function of time delay. One can immediately see that close to zero time delay the emission is relatively 
isotropic (the red color is uniformly distributed over all emission angles), while at later time delays the signal is peaked at $0^\circ$ and 
$180^\circ$, corresponding to emission along the laser polarization.  
      
Thus the angular distribution of photoemission significantly changes as a function of the time delay between the XUV pulse exciting the
Rydberg state and the IR pulse ionizing it. This observation can only be explained by the presence of two overlapping electronic states with
different lifetimes, since isolated electronic states (and hence IR-induced photoemission) cannot change with time. To confirm
this observation and extract parameters of the two contributing states we decompose the data matrix in Fig. \ref{fig:angmap}a) using the method
of singular value decomposition \cite{Golub1970, Yamaguchi1998}. 
This method leads to decomposition of the matrix into principal components.
For the case of two electronic states with different angular emission characteristics and different lifetimes we expect to observe a decomposition
of the data into two dominant contributions. This is indeed the case as can be seen in Fig. \ref{fig:angmap}b), which shows 
the data matrix composed of these
two dominant components with all other contributions removed. Clearly this map reproduces all features of the experimental data in part a, while
simultaneously strongly suppressing the noise. 

The singular value decomposition thus allows us to extract angular distributions and decay times of the two electronic states contributing to
the resonance at 17.33~eV \cite{Golub1970, Yamaguchi1998}. Exponential decay fits convoluted with a Gaussian cross correlation function
(FWHM of 36~fs) yield the lifetimes of $20\pm5$~fs and $9\pm3$~fs for the two contributions. The errors represent the statistical uncertainties
of the fit. Systematic uncertainties arising from the delay step size and the exact shape of the cross correlation function are relatively small for
the longer timescale, but are significant for the short one substantially extending the lower boundary of uncertainty. 
We therefore assume it to be less than 10~fs in the following text.
The timescale of 20~fs of the long-lived component is somewhat larger than the value of $14\pm1$~fs reported in Ref. \cite{Eckstein2016}. These values
are however consistent, because the data analysis used in Ref. \cite{Eckstein2016} (integration in a narrow angular range) does not fully remove the
contribution of the shorter timescale making the resulting value somewhat smaller (14~fs). The present analysis using the singular value decomposition is free from such contributions. 

\begin{figure}[htbp!]
  \centering
    \includegraphics[width=0.85\columnwidth]{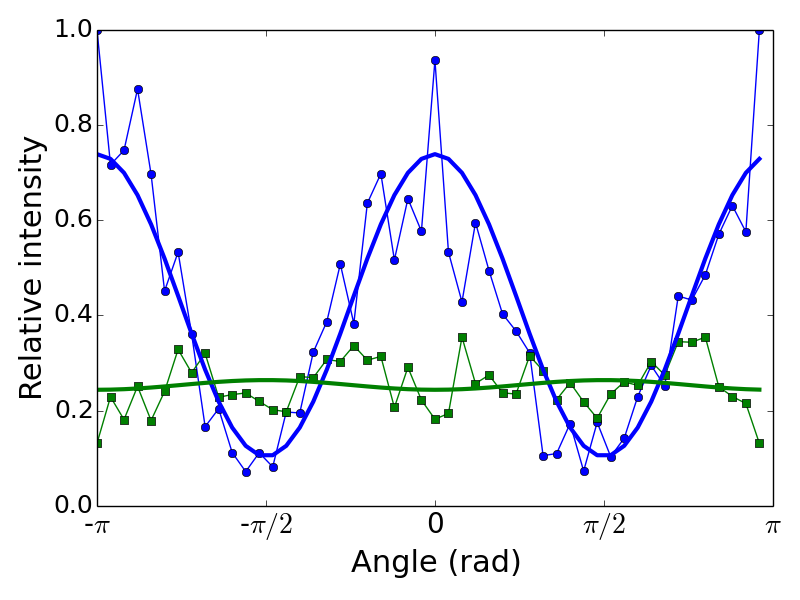}
      \caption{\label{fig:angtime} Decay-associated angular distributions of the two dominant components of the singular value decomposition, 
	  presented in Fig. \ref{fig:angmap}. Symbols are data points extracted in the SVD analysis. The two distributions correspond to $20\pm5$~fs 
	  decay (blue circles) and a decay shorter than 10~fs (green squares). Thick lines are fits of the asymmetry parameters yielding $\beta$ 
	  values of $1.3\pm0.2$ and $0.05\pm0.06$, respectively. }
\end{figure}

The decay-associated angular distributions of the long and short components are plotted in Fig. \ref{fig:angtime} as blue circles and green squares, 
respectively. 
The component with the longer lifetime peaks at $0^\circ$ and $180^\circ$. A least square fit to this distribution, shown as a blue line,
yields an asymmetry parameter of $1.35\pm0.15$. The short-lived component has a relatively uniform angular 
distribution (with an asymmetry parameter $\beta \approx 0$). We thus extract the parameters of the two electronic states contributing to the
resonance at 17.33~eV, which will be discussed later in the text in the context of the interference stabilization theory developed in the following
section.

\section{Theory of interference stabilization in autoionization}

The observation of two overlapping resonances with substantially different lifetimes immediately reminds of the interference stabilization theory
(or interference narrowing) developed
\cite{Heller1976, Feneuille1982, Liu1985, Fedorov1988, Fedorov1998} for atoms in static electric fields and intense laser fields. In the description of Ref. \cite{Fedorov1998} two Rydberg states of an atom are laser-coupled to the ionization continuum. When the coupling strength, expressed in units of energy,
exceeds the energy gap between the states, the mixing of the two states via the continuum leads to stabilization (increased lifetime) of one of them at
the expense of de-stabilization (decreased lifetime) of the other. This scenario is very general and can be applied to any type of
discrete-continuum couplings, i.e. also in the case of two overlapping autoionizing states of an atom or molecule interacting with the same ionization continuum. 
In the following we derive the theory of interference stabilization for the case of two autoionizing states coupled to two or more continua, which reflects the
realistic case of the $N_2$ molecule investigated in our experiment, where autoionization can proceed to the $X^2\Sigma_g^+$ and $A^2\Pi_u$ ionization
channels.

\subsection*{Stabilization effect}

We express the wavefunction of the molecule upon absorption of an XUV photon in terms of two discrete state wavefunctions $\phi_1$ and $\phi_2$ and two
continuum wavefunctions $\psi_{A}$ and $\psi_{X}$, with the total energies $E_1$, $E_2$, $E_{A}$ and $E_{X}$, respectively:

\begin{eqnarray}
\nonumber\Psi(t)&=&c_1(t)\phi_1e^{-iE_1t}+c_2(t)\phi_2e^{-iE_2 t}\\
&&+\int dE_{A} c_{A}(t)\psi_{A}e^{-iE_{A}t}+\int dE_{X} c_{X}(t)\psi_{X}e^{-iE_{X}t}
\end{eqnarray}

The labels $A$ and $X$ for the two continua are chosen to facilitate the subsequent comparison with the case of N$_2$. 
The matrix elements of the system Hamiltonian $H$ are given by the following set of equations:

\begin{eqnarray}
  \bra{\phi_n}H\ket{\phi_m} &=& E_n\delta_{nm}  \\
  \bra{\psi_{A}}H\ket{\psi'_{A}} &=& E_{A}\delta(E_{A}-E'_{A})  \\
  \bra{\psi_{X}}H\ket{\psi'_{X}} &=&E_{X}\delta(E_{X}-E'_{X})  \\
  \bra{\psi_{A}}H\ket{\psi_{X}} &=& 0  \\
  \bra{\phi_n}H\ket{\psi_{A}} &=& V_{An}  \\
  \bra{\phi_n}H\ket{\psi_{X}} &=& V_{Xn}  
  \end{eqnarray}

Here $\delta$ denotes the Dirac delta-function and $V_{An}, V_{Xn}$ are the terms describing the electron-electron interaction, which leads to autoionization \cite{Fano1961}.
Substituting the wavefunction $\Psi$ into the time-dependent Schr\"odinger equation and taking the projection of the total wavefunction $\Psi$ onto the wavefunctions of the individual states one arrives at the following system of coupled differential equations for state population coefficients $c(t)$:

\begin{eqnarray}
i\dot{c_1}(t)=\int dE_{A} c_{A}(t)e^{-i(E_{A}-E_1)t}V_{A1}+\int dE_{X} c_{X}(t)e^{-i(E_{X}-E_1)t}V_{X1}&\label{eq:systemstart} \\
i\dot{c_2}(t)=\int dE_{A} c_{A}(t)e^{-i(E_{A}-E_2)t}V_{A2}+\int dE_{X} c_{X}(t)e^{-i(E_{X}-E_2)t}V_{X2} &\\
i\dot{c_{A}}(t)=c_1(t)e^{-i(E_1-E_{A})t}V_{A1}+c_2(t)e^{-i(E_2-E_{A})t}V_{A2} &\\
i\dot{c_{X}}(t)=c_1(t)e^{-i(E_1-E_{X})t}V_{X1}+c_2(t)e^{-i(E_2-E_{X})t}V_{X2}&
\end{eqnarray}

As one can see from these equations, the coupling arises because both $\phi_1$ and $\phi_2$ couple to the same continuum states $\psi_{A}$ and $\psi_{X}$.
Following the treatment of Fedorov \cite{Fedorov1998} one can formally integrate equations for the continuum coefficients and substitute them into the
expressions for discrete states. Assuming that continuum coefficients $c_{A}$, $c_{X}$ and transition matrix elements $V_{An}$, $V_{Xn}$ are
sufficiently slow functions of the energy in the continuum channel ($E_{A}$ and $E_{X}$), the following system of differential equations is obtained:

\begin{eqnarray}
\label{eq:systemend}
i\dot{c_1}(t)&=&-2\pi i \left[\right.c_1(t)(|V_{A1}|^2+|V_{X1}|^2)+\nonumber\\
             & & c_2(t) e^{i(E_1-E_2)t}(V_{A1}V_{A2}+V_{X1}V_{X2})\left.\right]\\
i\dot{c_2}(t)&=&-2\pi i \left[\right.c_2(t)(|V_{A2}|^2+|V_{X2}|^2)+\nonumber\\
             & & c_1(t) e^{i(E_2-E_1)t}(V_{A1}V_{A2}+V_{X1}V_{X2})\left.\right]
\end{eqnarray}

\begin{figure}[ht!]
  \centering
    \includegraphics[width=\columnwidth]{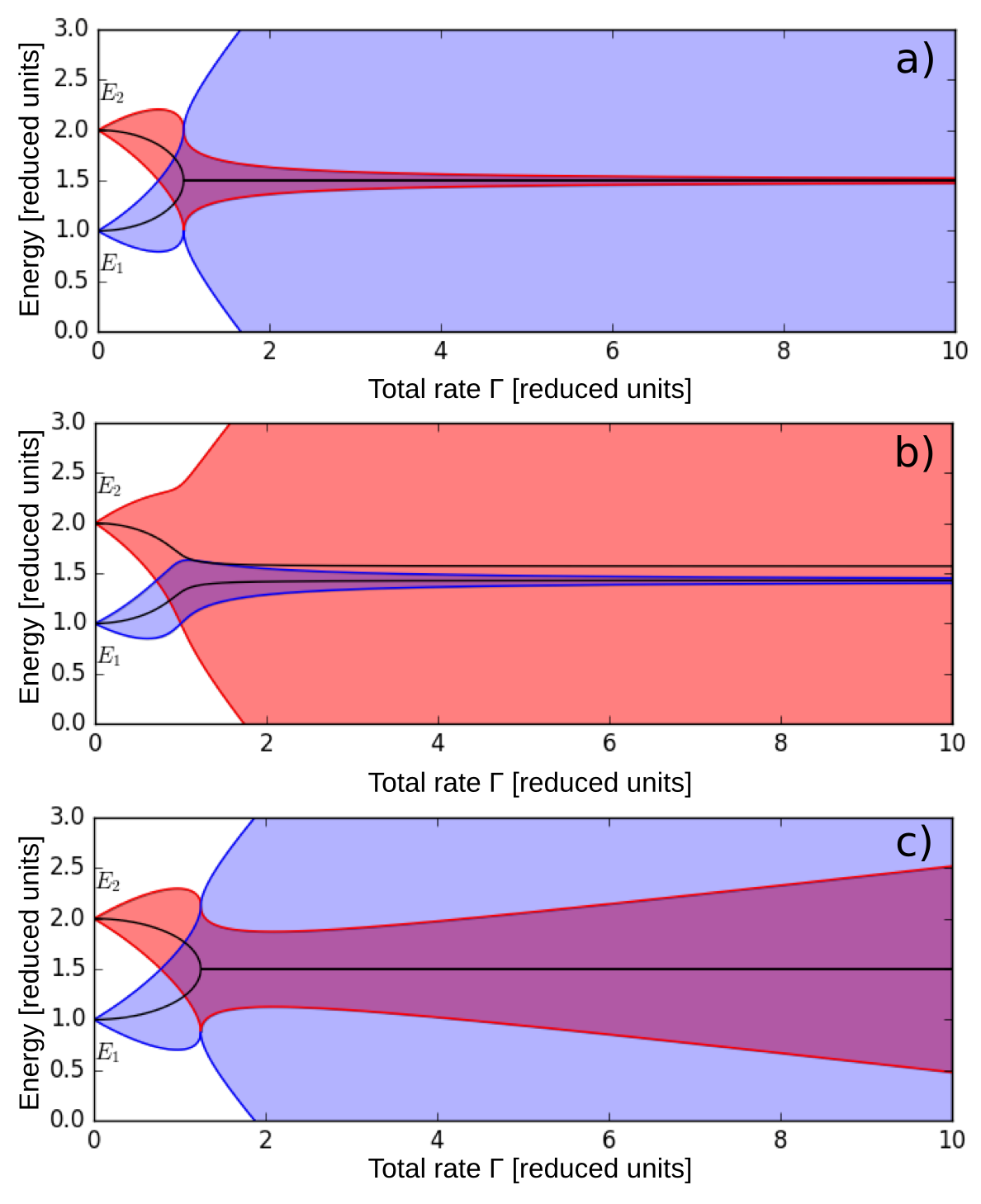}
      \caption{\label{fig:gammarates} Resonance positions $Re(\gamma_\pm)$ (black) and effective resonance boundaries $Re(\gamma_\pm) \pm Im(\gamma_\pm)$ 
       (red and blue) for three sets of parameters describing the discrete-continuum interactions, plotted against the total rate $\Gamma$ given in reduced units.
      a) Balanced case: the two discrete states have identical coupling strengths to the two continua, i.e. $V_{A1}=V_{A2}$ and $V_{X1}=V_{X2}$. b) Partially balanced case: the coupling strengths of both discrete states to the two continua are balanced ($V_{A1}=V_{X1}$ and $V_{A2}=V_{X2}$), but the 
      resonance 1 is coupled to the continua weaker than the resonance 2 ($\Gamma_1=0.75\Gamma_2$). c) Imbalanced cased: the coupling constants of
      two discrete states are different for each continuum: $\Gamma_{A1}=0.7\Gamma,~\Gamma_{X1}=0.3\Gamma,~\Gamma_{A2}=0.3\Gamma,~\Gamma_{X2}=0.7\Gamma$.}
\end{figure}

To compare with Ref. \cite{Fedorov1998} we may choose equal coupling strengths for both discrete states: $V_{A1}\simeq V_{A2}$ ($V_{X1}\simeq V_{X2}$) 
and define the decay rates $\Gamma_A\equiv 2\pi |V_{An}|^2$ and $\Gamma_X\equiv 2\pi |V_{Xn}|^2$. Looking for a non-trivial solution of the form 
$c_n(t)=f_n e^{i(E_n-\gamma)t}$ a quadratic equation for the quasi-energy $\gamma$ is obtained, which has two solutions:

\begin{equation}
\gamma_{\pm}=\frac{1}{2}\left[(E_1+E_2)-i(\Gamma_A+\Gamma_X)\pm\sqrt{(E_1-E_2)^2-(\Gamma_A+\Gamma_X)^2}\right]
\label{eq:quasienergy1}
\end{equation}

This is equivalent to the Eq. (8.1.21) of Ref. \cite{Fedorov1998} with the total decay rate of each discrete state equal to the sum $\Gamma=\Gamma_A+\Gamma_X$.
The resonance widths of the two interacting states are given by twice the imaginary part of the two quasi-energies $2 Im(\gamma_\pm)$
and the energy positions of the resonances are given by the real part $Re(\gamma_\pm)$. Fig. \ref{fig:gammarates}a) shows both the positions and the widths
of the resonances as a function of the total rate $\Gamma$. The positions are shown as black lines and the widths are represented by shaded areas enclosed by
effective resonance boundaries given by $Re(\gamma_\pm) \pm Im(\gamma_\pm)$. The energies and widths are given in reduced units chosen to yield the energy
spacing of the states at $\Gamma=0$ equal to 1.
One can see that at a critical value of $\Gamma=E_2-E_1$
the resonance positions merge and from there on the width of one of the resonances continues to grow, while the width of the other, remarkably,
shrinks, indicating an increase of the lifetime with increasing coupling strength. 
Therefore one of the state becomes more and more stabilized. The critical
rate $\Gamma$ corresponds to a change of the sign of the expression under the square root in Eq. \ref{eq:quasienergy1}.
For the case of molecular autoionization considered here the continuum couplings are given by electron-electron interactions and cannot be changed
experimentally. Therefore the case of the molecule would correspond to a particular value of $\Gamma$ given by the electron-
electron interactions.

Generalizing the result of Ref. \cite{Fedorov1998}, the couplings of two discrete states to the two continua are not identical and, introducing decay rates 
$\Gamma_{An}=2 \pi |V_{An}|^2$, $\Gamma_{Xn}=2 \pi |V_{Xn}|^2$ and carrying out the same procedure as outlined above we arrive at the following expression for the quasi-energy:

\begin{equation}
\gamma_{\pm}=\frac{1}{2}\left[(E_1+E_2)-i(\Gamma_1+\Gamma_2)\pm\sqrt{\Delta}\right],
\label{eq:quasienergy2}
\end{equation}

where $\Gamma_1=\Gamma_{A1}+\Gamma_{X1}$, $\Gamma_2=\Gamma_{A2}+\Gamma_{X2}$ and $\Delta$ is given by: 

\begin{equation}
\Delta=(E_1-E_2)^2+i(\Gamma_2-\Gamma_1)-4(\sqrt{\Gamma_{A1}\Gamma_{A2}}+\sqrt{\Gamma_{X1}\Gamma_{X2}})^2
\end{equation}

One can immediately see that for the case $\Gamma_1=\Gamma_2$ the value of $\sqrt{\Delta}$ is always real and the system behaves similar to
that of Ref. \cite{Fedorov1998} when all continuum couplings are uniformly increased leading to stabilization of one of the resonances.
In the general case, however, the couplings of the two resonances to the two continua are different and this leads to a modification of the system behavior.
If $\Gamma_1 \neq \Gamma_2$, but the coupling strengths of the discrete states to the two continua remain \textbf{balanced}, i.e. 
$\Gamma_{A1}=\Gamma_{X1}$ and $\Gamma_{A2}=\Gamma_{X2}$, the resonance positions do not merge anymore, but the stabilization effect is still observed,
as can be seen in Fig. \ref{fig:gammarates}b). The positions of the two resonances do move closer together and the resonance with the initially 
weaker coupling to the continua (i.e. resonance 1 for $\Gamma_2>\Gamma_1$) becomes stabilized.
 
Finally, an interesting effect is observed for the general case of \textbf{imbalanced} coupling strengths, i.e. when either $\Gamma_{A1}\neq\Gamma_{X1}$
or $\Gamma_{A2}\neq\Gamma_{X2}$ or both. In this case, for a certain degree of imbalance the weaker coupled state first becomes partially stabilized, but then passes through a minimum of the width and diverges (see Fig. \ref{fig:gammarates}c)). That is, there is an optimum stabilization
condition and a minimum non-infinite lifetime a state can achieve. For an extreme imbalance the stabilization effect can disappear altogether. An example is a limiting case of two discrete resonances coupled to different continua, i.e. $\Gamma_{A2}=0$ and $\Gamma_{X1}=0$. 

\subsection*{Spectral lineshapes}

Equation (\ref{eq:quasienergy2}) for the quasi-energies can be used to derive the temporal dynamics of the system of two autoionizing resonances and obtain the time-dependent dipole response to the XUV pulse. For this purpose the system of equations (\ref{eq:systemstart})-(\ref{eq:systemend}) is modified to include the ultrafast excitation ($\delta$-like) by a XUV pulse at time $t=0$:

\begin{eqnarray}
i\dot{c}_i = -d_{gi}\delta(t) + \sum_{J}\int dE_J c_{J}(t) e^{-i(E_J-E_i)t}V_{iJ}\\
i\dot{c}_{J} = -d_{gJ}\delta(t) + \sum_{l} c_l(t)e^{-i(E_l-E_J)t}V_{lJ}
\end{eqnarray}

where we generalize the treatment to an arbitrary number of continua denoted by the subscript $J$ (i.e. $J=\{A,X\}$ in the previous section). 
The parameters $d_{gi}$ and $d_{gJ}$ describe
the transition dipole moments coupling the neutral ground state $\ket{g}$ and the discrete and continuum states respectively. As before, they are considered
to be independent of the final energy in the continuum $E_J$. Following the same procedure as before, the solutions for the populations of discrete states
can be cast in the following form: 

\begin{equation}
c_i(t)=\sum_{\alpha=\pm}C_{\alpha}A_{i}^{\alpha}e^{-i\gamma_{\alpha}t}e^{iE_it},
\end{equation}

where the coefficients $A_{i}^{\pm}$ describe the mixing of the discrete states via interference stabilization:

\begin{eqnarray}
A_1^{\pm}&=&1\\
A_2^{\pm}&=&\frac{i}{2}\left(\frac{(E_2-E_1)-i(\Gamma_{2}-\Gamma_{1})\pm\sqrt{\Delta}}{\sum_{J}\sqrt{\Gamma_{J1}\Gamma_{J2}}}\right)
\end{eqnarray}

and the coefficients $C_{\pm}$ are set to fulfill the initial conditions given by the XUV pulse:

\begin{eqnarray}
C_{+}&=&\frac{2\pi i}{A_2^{-}-A_{2}^{+}}\sum_{J}d_{gJ}\left[V_{1J}(q_1-i)A_{2}^{-}-V_{2J}(q_2-i)\right]\\
C_{-}&=&\frac{2\pi i}{A_2^{-}-A_{2}^{+}}\sum_{J}d_{gJ}\left[V_{2J}(q_2-i)-V_{1J}(q_1-i)A_{2}^{+}\right]
\end{eqnarray}

where we set the Fano indices of the re-normalized discrete states as:

\begin{equation}
q_i=\frac{d_{gi}}{2\pi\sum_{J}d_{gJ}V_{iJ}}.\label{eq:qfano}
\end{equation}

The time-dependent dipole response is then given by the following expression:

\begin{eqnarray}
&&d(t) \equiv \bra{g}\hat{H}_{XUV}\ket{\Psi(t)}=\\
&&\sum_{J}\left[\sum_{i,\alpha}C_{\alpha}A_{i}^{\alpha}e^{-i\gamma_{\alpha}t}2\pi d_{gJ}V_{Ji}(q_i-i)+2i\pi|d_{gJ}|^2\delta(t)\right]\Theta(t)\nonumber
\end{eqnarray}

The energy-domain absorption profile can be obtained by carrying out the Fourier transform of the time-dependent dipole response $d(t)$ and taking the imaginary part of the resulting energy-domain complex response function $Im[d(E)]$ \cite{Ott2013}. This procedure is implemented in the next section to compare the
formalism developed here with the experimental data.

\section{Discussion}

The theoretical treatment of interference stabilization in the presence of several continua given above can now be used to analyze the Fano resonance in $N_{2}$
at 17.33~eV, using the current experimental observations as well as the most detailed theoretical calculations available of Raoult et al \cite{Raoult1983}.
According to the calculations of Raoult, two Rydberg states may contribute to this resonance: the {\resd} and the {\ress} states. Based on 
calculated absorption strengths the authors assign the dominant contribution to the
{\resd} state, which is coupled to two continua with the N$_2^+$ ion in 
the $X^2\Sigma_g^+$ and $A^2\Pi_u$ states. Since the calculations of Ref. \cite{Raoult1983} are performed in a space-fixed molecular frame, the full symmetry
dictated by the electronic wave function of the excited Rydberg state should be preserved during autoionization and therefore the
{\resd} state is coupled to $X\Pi$ and $A\Pi$ continua, where $\Pi$ describes the product symmetry of the ion core and the outgoing electron. Similarly, the
{\ress} Rydberg state in their calculations is coupled to the $X\Sigma$ and $A\Sigma$ continua. The coupling constants and relative populations of the discrete and continuum states can be extracted from Fig. 5 of Ref. \cite{Raoult1983} by fitting Fano profiles to the theoretical curves. The parameters of the Fano profiles obtained this way are given in Table \ref{tab:Raoult_fit}.

\begin{table}[htbp!]
\small
  \caption{\ Parameters of Fano profiles extracted from Fig. 5 of Ref. \cite{Raoult1983} for two discrete states coupled to four possible
  ionization continua. }
  \label{tab:Raoult_fit}
  \centering
  \begin{tabular*}{0.3\textwidth}{@{\extracolsep{\fill}}lll}
    {\resd} & X$\Pi$ & A$\Pi$ \\
    \hline
    ~$E_0$~[eV]	   & 17.35 & 17.35 \\
    ~$\Gamma$~[meV] & 63 & 74 \\
    ~q             & 1.18  & 0.37  \\
    \\
    {\ress} & X$\Sigma$ & A$\Sigma$  \\
    \hline
    ~$E_0$~[eV]	   & 17.405 & 17.405 \\
    ~$\Gamma$~[meV] & 63  & 88 \\
    ~q             & -1.08  & 4.87  \\
  \end{tabular*}
\end{table}

The treatment of Ref. \cite{Raoult1983} does not include the effect of interference stabilization, because in the fixed molecular frame 
the two contributing discrete states are coupled to ionization continua of different symmetry, as can be seen in the Table \ref{tab:Raoult_fit}. 
The resonance widths $\Gamma$ given in the table are all rather similar, ranging from 63 to 88~meV. This, however, contradicts our experimental observations 
which clearly show that the two states have substantially different lifetimes and, therefore, must have different spectral widths. This controversy can be resolved
if we recall that the experiments are not carried out in a fixed molecular frame. In the gas phase, nitrogen molecules can freely rotate and the autoionizing transitions may involve changes in the rotational state of the molecule. Such rotational transitions lead to a coupling
of states with different symmetries, because the angular momentum of the total wave function includes the angular momentum of the changing rotational state.
Indeed in 1993 Huber et al \cite{Huber1993} re-investigated the Hopfield series both at room temperature and in rotationally cold supersonic
jets and found significant distortions of the line profiles induced by rotation of the molecules. They pointed out that for high rotational temperatures the effect of ``l-uncoupling''
should be taken into account, that is the coupling of the electronic orbital angular momentum to the molecular axis becomes weaker than the rotational couplings. In molecular spectroscopy this corresponds to transition from the Hund's case (b) (fixed frame) to the Hund's case (d) (rotational couplings). In the latter case the projection $\Lambda$ of the electronic orbital angular momentum on the molecular axis is not a good quantum number, the Rydberg states acquire mixed $\Sigma/\Pi$ character and can interact with each other via ionization continuum.

\begin{figure}[htbp!]
  \centering
    \includegraphics[width=\columnwidth]{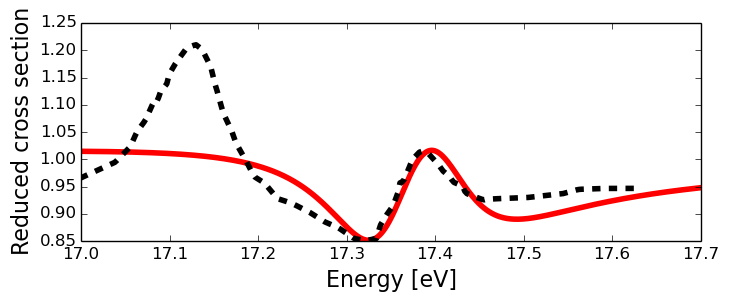}
      \caption{\label{fig:nointerf} Calculated XUV absorption profile of the resonance at 17.33~eV taken from Ref \cite{Raoult1983} (black dashed line) and
               calculated using the theory developed in the present work not including the stabilization effect. The resonance at 17.14~eV is not included in
               our treatment.}
\end{figure}

We therefore propose that rotational couplings relax the symmetry constraints in autoionizing transitions and allow for electron-electron interactions between discrete states and continua of different symmetry. This in turn leads to interactions between the two resonances affected by the effect of interference stabilization. 
To test this assumption we implement the model described in the previous section for two discrete states (state 1, {\resd}, and state 2, {\ress}) 
coupled to four continua ($X\Pi,~A\Pi,~X\Sigma$ and $A\Sigma$). We set the coupling strengths for the symmetry-allowed transitions based on the values of the resonance widths $\Gamma$ given in Table \ref{tab:Raoult_fit}. The Fano indices q given in the 
Table are used to set the relative populations of the discrete states and the continua induced by absorption of the XUV. The only remaining parameter, i.e. the
coupling strengths of the two discrete states to the neutral ground state $d_{g1}$ and $d_{g2}$, are set to $d_{g2}/d_{g1}=3$ to match the XUV absorption 
spectrum calculated by Raoult at al (Fig. 4 of Ref. \cite{Raoult1983}). 
The results of the calculation corresponding to the case of Ref. \cite{Raoult1983}, i.e. without symmetry-changing transitions and thus, without interference stabilization, are shown in Fig. \ref{fig:nointerf} together with the calculated absorption spectrum of Ref. \cite{Raoult1983}. The general shape of the resonance at 17.33~eV is well reproduced with all differences attributable to the resonance at 17.14~eV, which is omitted in our calculations. 

\begin{figure}[ht!]
  \centering
    \includegraphics[width=\columnwidth]{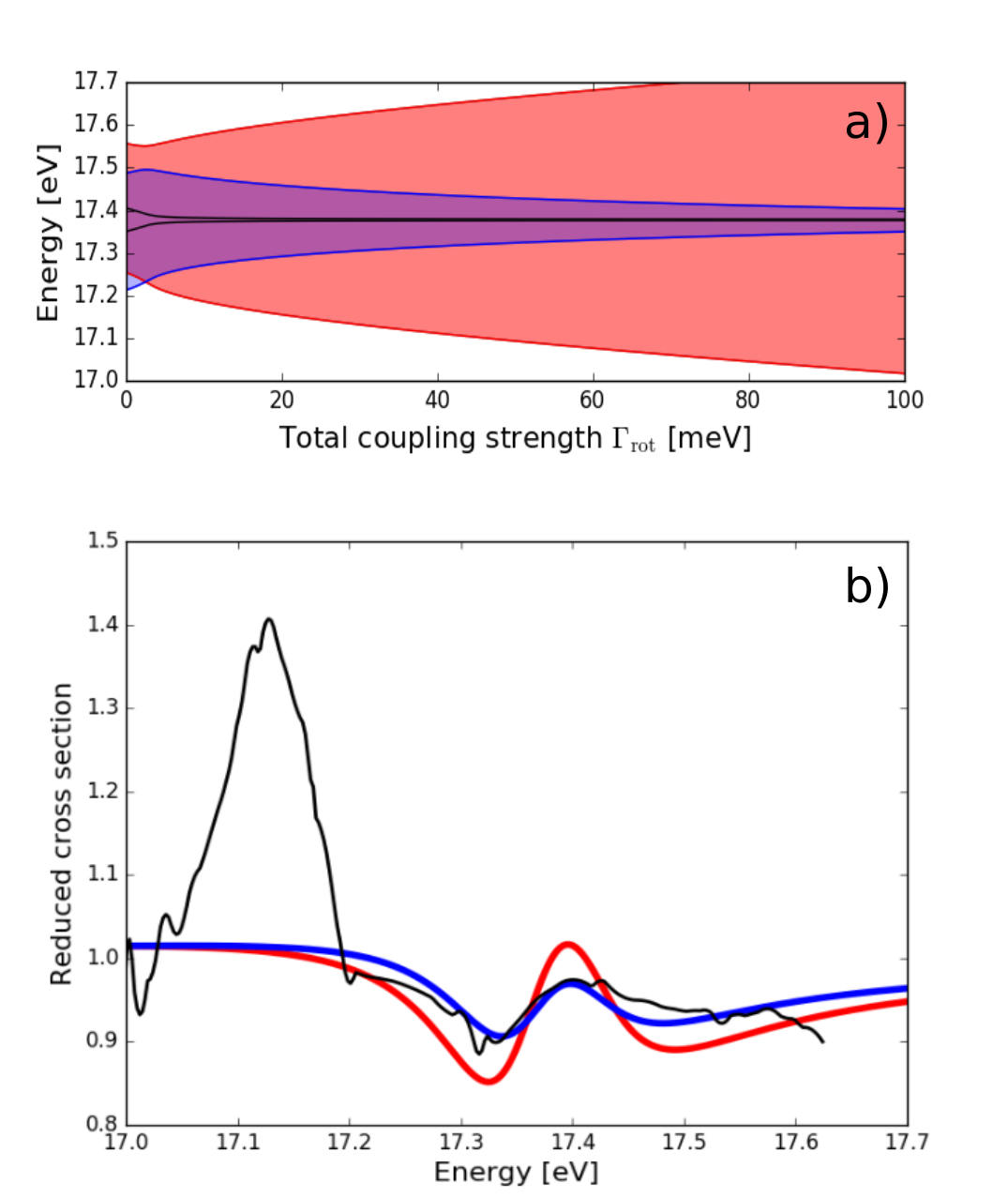}
      \caption{\label{fig:interf} a) Quasi-energy diagram similar to those of Fig. \ref{fig:gammarates} demonstrating the effect of interference stabilization
               induced by the symmetry-changing rotational couplings $\Gamma_\mathrm{rot}$. The case of $\Gamma_\mathrm{rot}=0$ corresponds to the calculations 
of Ref. \cite{Raoult1983}. b) XUV absorption profile calculated with (thick blue line) and without (thick red line) the effect of interference stabilization compared to the experimental spectrum of Morin et al \cite{Morin1983} (black line). This plot demonstrates how the inclusion of interference stabilization can largely remove the discrepancies observed in the fixed molecular frame calculations of Raoult et al \cite{Raoult1983}.}
\end{figure}

We can now model the effects induced by rotations. To the best of our knowledge there is no information in the literature on rotational coupling strengths for these short-lived 
Rydberg states. To test our assumption we vary the values of the corresponding symmetry-changing matrix elements, describing coupling between the $3d\pi_g$ Rydberg state and the $X\Sigma$ and $A\Sigma$ continuum states, and between the $4s\sigma_g$ Rydberg state and the $X\Pi$ and $A\Pi$ continuum states. For simplicity 
we take all couplings, initially forbidden by symmetry, to be equal and set them to values ranging from 0 to 25~meV, which corresponds to maximum total
rotational coupling of 100~meV. For a comparison, remember that the symmetry-allowed couplings in the Table \ref{tab:Raoult_fit} are all in the range of
63 to 88~meV.
The quasi-energy diagram for this range of total rotational couplings $\Gamma_\mathrm{rot}$ is shown in Fig. \ref{fig:interf}a). Note, that only
rotational couplings are varied in the plot and therefore at the $\Gamma_\mathrm{rot}=0$ the widths of the states are given by sums of the values in Table \ref{tab:Raoult_fit}.
One can see that already for total rotational coupling on the order of 5~meV the positions of the two discrete states (shown by black lines in the plot) approach each
other, leading to substantial modification of the absorption profile. The stabilization effect also sets in at very small values of total coupling strength and for $\Gamma_\mathrm{rot}=20$~meV the ratio of the lifetimes (or spectral widths) exceeds 2. For much larger values of $\Gamma_\mathrm{rot}$ (not shown in the plot) the resonance is destabilized as predicted in the general imbalanced case of Fig. \ref{fig:gammarates}c). In Fig. \ref{fig:interf}b) XUV absorption profiles calculated with the total rotational couplings of 0~meV and 1~meV are compared with the experimental data of Morin et al \cite{Morin1983}. The plot shows that even very small couplings can explain the discrepancies observed in the calculations of Raoult et al \cite{Raoult1983}.

The relative contributions of the resonance positions and resonance width to the combined XUV absorption profile sensitively depend on the specific values 
of the couplings to different continua. An analysis of the photoelectron angular distributions in Fig. \ref{fig:angtime} shows that the long-lived component has a relatively large positive asymmetry parameter $\beta=1.3\pm0.15$. For linearly polarized light such a distribution corresponds to one-photon ionization of a state 
with a dominant s-type contribution. Therefore the angular distributions observed in the experiment are consistent with stabilization of the
{\ress} Rydberg state induced by rotational couplings of angular momenta. Considering the results of the theoretical section presented in Fig. \ref{fig:gammarates}b) we can therefore conclude, that the {\ress} state is the one with a weaker coupling to the continuum. This conclusion is indeed logical
if one considers the physics of couplings induced by rotations. For a perfectly spherically symmetric Rydberg state (s-state) rotations of the molecular axis do not lead to changes in the electron density distributions and therefore one may consider this state decoupled from the rotational motions. The d-state, on the contrary,
has a symmetry axis along the molecular axis and thus should also rotate together with the molecule. Therefore it is reasonable to expect the 
rotationally-induced couplings of the {\resd} state to be larger than those of the {\ress} state. This in turn means that the weaker coupled {\ress} state
is expected to be stabilized, which is what we observe in the experiment (see Fig. \ref{fig:angtime}).

The model formulated above includes rotational couplings as a variable parameter. A first principles calculation of these couplings would require quite sophisticated methods, because
the effect of l-uncoupling essentialy corresponds to the breakdown of the Born-Oppenheimer approximation and requires a special treatment. Additionally, Rydberg states discussed here have lifetimes shorter than the rotational period of the ion core, which means that the rotational angular momentum and its projection are not good quantum numbers in the excited state. Another approach could be based on the MQDT theory, such as that used by Raoult et al \cite{Raoult1983}. The rotational couplings can possibly be inferred from high resolution spectroscopic experiments on high members of the Rydberg series, which have longer lifetimes and permit resolution of the rotational structure.

The interference stabilization mechanism presented here is general and may play a role in many phenomena where several discrete states are
sufficiently strongly coupled to the same continuum. Such situations can arise, for example, in autoionization or Auger decays of 
atoms and molecules and lead to very complex resonance structures, which cannot be easily assigned based on static spectroscopic methods. In these 
situations time-dependent techniques prove to be very useful, as demonstrated here, and could help to resolve many challenging problems in 
spectroscopy.
 
\section{Conclusions}

This article presents results of the time-, energy- and angular-resolved photoelectron spectroscopy of a single autoionizing 
resonance of N$_2$ at 17.33~eV. The experimental photoelectron spectra reveal a transient feature corresponding to ionization of autoionizing Rydberg states
excited by a short XUV pulse. Analysis of angular distributions reveals two dominant contributions to the photoelectron spectra
corresponding to two transient electronic states. One state has a relatively long lifetime of $20\pm5$~fs and a peaked
angular distributions revealing its s-like character. The other state has a very short lifetime ($<10$~fs) and a 
uniform angular distribution. Such combination of two overlapping resonances with long and short lifetimes has been previously predicted and 
observed for the case of laser-induced interaction of Rydberg states with an ionization continuum and 
is referred to as the effect of interference stabilization. Here we develop a formalism of interference stabilization for the case of autoionizing molecular resonances and apply it to describe the window resonance investigated in our experiments. We propose, that rotations of the molecule induce the otherwise symmetry-forbidden coupling between the two contributing Rydberg states, the {\resd} and {\ress} states attached to the $B^2\Sigma_u^+$ ionic core, and 
lead to a stabilization of one at the expense of the other. Using published values for the coupling strengths of the two contributing Rydberg states, 
and adjusting the strength of the unknown rotational couplings we are able to remove previously unexplained discrepancies between the calculated and experimental 
XUV absorption spectra. This observation suggests that even very weak couplings of electronic states can be amplified by the effect of interference 
stabilization and significantly modify congested XUV spectra. Our results both demonstrate the advantages of time-resolved methods for the analysis of complex 
overlapping autoionizing resonances and provide a general theoretical treatment of interference stabilization in molecular autoionization, 
which can be applied to a large variety of problems in autoionization, Auger decay and similar phenomena in atoms and molecules.

\section{Acknowledgments}

We thank Iason Katechis for help with the SVD analysis.

\bibliography{N2_Fano_Faraday}
\bibliographystyle{rsc}
\end{document}